\documentclass[a4paper,11pt]{article}
\usepackage{amsmath}
\usepackage{graphicx}
\usepackage[left=2.2cm, right=2.2cm, top=2cm]{geometry}
\usepackage{algorithm}
\usepackage{algpseudocode}
\usepackage{subfigure}
\usepackage{color}
\usepackage{multirow}

\numberwithin{figure}{section}

\begin{document}
\title{Encoding and Decoding Algorithms for Arbitrary Dimensional Hilbert Order}
\author{Hui Liu, Tao Cui, Wei Leng and Linbo Zhang}
\date{}
\maketitle

\begin{abstract}
Hilbert order is widely applied in many areas. However, most of the algorithms are confined to low
dimensional cases. In this paper, algorithms for encoding and decoding arbitrary dimensional
Hilbert order are presented. Eight algorithms are proposed. Four algorithms are based on arithmetic operations
and the other four algorithms are based on bit operations. For the algorithms complexities, four of them are
linear and the other four are constant for given inputs. In the end of the paper, algorithms for two
dimensional Hilbert order are presented to demonstrate the usage of the algorithms introduced.\bigskip
\end{abstract}

\section{Introduction}

Space-filling curves were proposed by Peano in 1890 and popularized
by Hilbert later. These curves introduce maps between one dimensional domain and multiple dimensional domain,
which also introduce order when considering in one dimensional space.
Hilbert space-filling curve is one famous space-filling curve, which is also called Hilbert
curve. This kind of curve has many important characteristics, such as locality,
clustering and self-similarity. Hilbert curve (order) has been applied in many areas,
including image storing, database indexing, data compression and dynamic load balancing.
For parallel computing, Hilbert order method is one of the most important
geometry-based partitioning methods, which was implemented by the Zoltan \cite{Zoltan}, one of the most
well-known dynamic load balancing package developed by Sandia National Laboratories. It was also implemented
by PHG (Parallel Hierarchical Grid) \cite{phg}, and it serves as the default dynamic load balancing strategy.
The Hilbert order now are widely applied by many parallel applications \cite{sfc-tr-03, hliu}.
Details of space-filling curves can be found in \cite{Hans}.

A $n$-dimensional Hilbert curve introduces a one-to-one mapping between $n$-dimensional space and one-dimensional space.
The mapping from $n$-dimensional space to one-dimensional space is called encoding, while the inverse mapping is called decoding, which maps an integer to a coordinate in $n$-dimensional space.
Algorithms for computing Hilbert curve/order in two and three dimensional spaces have been proposed in
many literatures, which can be classified into recursive algorithms
\cite{butz, gold, witten, cole} and iterative algorithms \cite{griff, sfc-tr-03, xliu2, xliu, fisher, ningtao}.
Iterative algorithms, especially the table-driven algorithms \cite{griff,
sfc-tr-03}, are usually much faster than recursive algorithms. In general, the complexities of these algorithms are $O(m)$, where $m$ is
the level of the Hilbert curve. For two dimensional space, Chen \emph{et al} \cite{ningtao} proposed an algorithm of $O(r)$ complexity,
where $r$ is defined as $r=\log_2(\max(x,y))+1$, $r \le m$ and is independent of the level $m$.
The algorithm is faster when $m$ is much larger than $r$. The same idea was also applied to three dimensional space \cite{hliu}.
For higher dimensional spaces, relative little work has been done due to the complexity of
Hilbert curve. Recently, Kamata \textit{et al} presented a representative $n$-dimensional Hilbert mapping algorithm
\cite{kamata} and Li \textit{et al} introduced algorithms for analyzing the properties of
$n$-dimensional Hilbert curve \cite{chenyang}.

In this paper, we present our work on developing algorithms for encoding and decoding arbitrary high dimensional
Hilbert order bases on \cite{chenyang}, and the authors introduced novel algorithms for
analyzing evolutive rules for high dimensional Hilbert curve. However, the decoding and encoding algorithms
for high dimensional spaces are still open problems.
Here the rules are studied firstly and some properties are deduced. Then the encoding and decoding algorithms
are proposed based on the properties deduced.
For the encoding problems, four algorithms are presented. Two algorithms are based on arithmetic operations,
which can be translated to bit operations naturally. Then two bit operation based algorithms are obtained.
The decoding processes are reverse processes of encoding, thus decoding algorithms are obtained similarly.
Four of our eight algorithms have linear complexity, O(m), where $m$ is the level of Hilbert order. The other
four algorithms have constant complexity for given inputs. In the end of the paper, the algorithms
are demonstrated in two dimensional space, and the results are discussed.

The layout of the paper is as follows. In \S \ref{sec-pre}, background and some notations are
introduced firstly and then algorithms proposed in \cite{chenyang} are analyzed in detail.
In \S \ref{sec-alg}, encoding and decoding algorithms are proposed.
In \S \ref{sec-dis}, a two dimensional case is employed to
illustrate our algorithms, and numerical experiments are performed to show the difference between
linear complexity algorithm and constant complexity algorithm.

\section{Preliminary}
\label{sec-pre}

Notations are introduced and then the algorithms proposed in \cite{chenyang} are studied in detail.
We should mention that all operations in this paper are performed on non-negative integers.

\subsection{Notations}
Let $n \ (n \ge 2)$ be the dimension of Hilbert curve
and $m$ be the level of Hilbert curve. Let $D_m$ be the coordinate set of the $m$-th level Hilbert curve, which
is defined as $D_{m}= \{(x_n,\cdots, x_2, x_1)| 0 \le x_i < 2^{m}, 1
\le i \le n\}$. $(x_n, \cdots, x_1) (\in D_m)$ is a coordinate of the Hilbert curve, and $x_i ( 1 \le i \le n)$ is
called the $i$-th component of the coordinate. Logical operation $\wedge$ for two coordinates is defined as
\begin{equation}
(x_n, \cdots, x_2, x_1)\wedge (y_n, \cdots, y_2, y_1) = (x_n\wedge
y_n, \cdots, x_2\wedge y_2, x_1\wedge y_1),
\end{equation}
where ${\wedge}$ is the regular exclusive or operation (xor).

Let $(a_1a_2 \cdots a_k)_{d}$ represent a number system, where $ 0 \le
a_i < d (1 \le i \le k)$, and $k$ can be any positive integer.
The number is binary number for $d=2$, and decimal number for
$d=10$. For any non-negative integer $j$, where $j = (a_1a_2\cdots a_k)_2$,
$Re_k$ is defined as
\begin{equation}
Re_k(j) = Re_k((a_1a_2\cdots a_k)_2) = (b_1b_2\cdots b_k)_2, b_i = 1
- a_i, 1 \le i \le k.
\end{equation}
The and, right and left shift operators are also introduced for
the case that $d$ equals to two, denoted by $\&$, $\gg$ and $\ll$, respectively.
Let us define,
\begin{equation}
p_n^{i}(a_1,a_2, \cdots, a_n) = (\sum\limits_{j = 1}^i {a_i}) ~mod~ 2 = (\sum\limits_{j = 1}^i {a_i}) ~\%~ 2,
\end{equation}
where $ a_i$ equals to 0 or 1$(1 \le i \le n)$. The $p_n^{i}(a_1,a_2, \cdots, a_n)$ equals to 1 or 0.
With the help of $p_i$, the function $f_n$ is defined as
\begin{equation}
\label{forward} f_n(a_1,a_2, \cdots ,a_n) = (b_1b_2\cdots b_n)_2 =
j, b_1 = a_1, b_i =\left\{\begin{array}{rl}
a_i, ~if ~p_n^{i-1} = 0  \\
1 - a_i,~if ~p_n^{i - 1} = 1
\end{array}\right. ,
\end{equation}
where $a_i$ equals to 0 or 1 and $j$ is a decimal number. The
function $f_n$ maps a vector $(a_1, a_2, \cdots, a_n)$ to a decimal number $j$.
Its inverse function $b_n$ is defined as
\begin{equation}
\label{backward} b_n(j) = b_n((a_1a_2 \cdots a_n)_2) =
(b_1,b_2,\cdots,b_n), b_1 = a_1, b_i =\left\{\begin{array}{rl}
a_i, ~if ~a_{i-1} = 0  \\
1 - a_i,~if ~a_{i - 1} = 1
\end{array}\right. ,
\end{equation}
where $j (0 \le j < 2^n)$ is a decimal number and $j = (a_1a_2 \cdots a_n)_2$.
The function maps a decimal number (scalar) to a vector.

\textbf{\emph{Remarks.}} In practice, for a fixed value $n$, $b_n$ and
$f_n$ can be calculated aforehand and saved in tables, and this
technique would speed up programs. $b_n$ can also be calculated in the
following way
\begin{equation}
b_n((a_1\cdots a_n)_2) = (a_1,\cdots, a_n) \wedge  (0,a_1,\cdots,
a_{n-1}).
\end{equation}

For the sake of completeness, concepts introduced in~\cite{chenyang} are borrowed here. An
$n$-dimensional Hilbert cell is a $1$-th level $n$-dimensional Hilbert curve. The $n$-dimensional Hilbert gene is a list of
coordinate transformation commands, which direct the generation of the $m$-th level Hilbert curve from $(m-1)$-th level Hilbert curve. The
coordinate transformation commands include two types: exchange
command and reverse command. These commands can be interpreted as reflection from some hyperplane, which
will be presented in the next section.
The Hilbert curve has direction, which has an entry and an exit. When the transformation commands are performed, the
coordinates of entry and exit may be changed. In \cite{chenyang}, $H_n^{i,0}$ and $H_n^{i, 1}$, where $ 0 \le i < 2^n$, were introduced
to represent the entry and exit for quadrant $i$. Algorithms were also
proposed to generate entry, exit and Hilbert gene list. Details can be found in \cite{chenyang}.

\subsection{Study of Evolutive Rules}
Now we analyze the properties of algorithms developed in \cite{chenyang}.
Hilbert cell was generated in recursively in \cite{chenyang}, in which if we wanted to
generate $n$ dimensional Hilbert cell,
the $n-1$ dimensional Hilbert cell should be generated aforehand.
By using the function $b_n$ instead, the $n$ dimensional Hilbert cell can be generated directly.

The Hilbert gene list \cite{chenyang} includes two types of commands: exchange command
and reverse command. Using mathematical techniques developed in this paper, each command can be
interpreted as reflection from some hyperplane. For exchange command, if the command is
performed between $x_i$ and $x_j$ $(i \ne j)$, the
hyperplane is $x_i - x_j = 0$, which means, for a point $P$, $(x_n, \cdots, x_i, \cdots, x_j,\cdots, x_1)$,
the coordinate of the new point $P'$ is $(x_n, \cdots, x_j, \cdots, x_i,\cdots, x_1)$.
This command just swaps $x_i$ and $x_j$. For reverse command, if the command is performed on
the $i$-th position, and if
the current level of Hilbert curve is $m$, the hyperplane is $x_i = \frac{2^m - 1}{2}$,
which means for any point $P$, $(x_n, \cdots, x_i, \cdots,
x_1)$, the coordinate of the new point $P'$ is $(x_n, \cdots, 2^m - 1 - x_i, \cdots, x_1)$. Here $(2^m - 1 - x_i)$ equals to $Re_m(x_i)$.
The reverse command changes $x_i$ only.

The algorithms which generateed gene list \cite{chenyang} can be rewritten as
\begin{equation}
\label{exchange} G_n^{i,0} = (b_n(0) \wedge b_n(2^n -
1))\wedge(H_n^{i,0} \wedge H_n^{i,1}),
 (0 \le i < 2^n)
\end{equation}

\begin{equation}
\label{reverse} G_n^{i,1} = b_n(0)\wedge H_n^{i,0}, ( 0 \le i <
2^n),
\end{equation}
where $G_n^{i,0}$ and $G_n^{i,1}$ represent the exchange command and the reverse
command respectively. According to the definition of $b_n$, $b_n(0)$ and $b_n(2^n -1)$ can be written as $(0,\cdots,0)$ and $(1,0, 0,
\cdots, 0)$. In this case, $b_n(0) \wedge b_n(2^n - 1)$ equals to $(1,0, 0, \cdots, 0)$. All the components of entry $H_n^{i,0}$
and exit $H_n^{i, 1}$ are the same except one, which means $G_n^{i,0}$ has no or only two components that
equal to 1. If the $i$-th and $j$-th components of $G_n^{i,0}$ are
1, then exchange command should be performed between $x_i$ and
$x_j$. The exchange command isn't performed if all components of $G_n^{i, 0}$ are 0. For $G_n^{i,1}$, if the $i$-th component is 1,
the reverse command should be performed in the $i$-th position. The executed order of reverse
commands doesn't affect the final result.

For quadrant $0$, $b_n(0)$, $b_n(1)$, $H_n^{0, 0}$ and $H_n^{0, 1}$ are $(0, \cdots, 0)$, $(0, \cdots, 0, 1)$, $(0, \cdots, 0)$ and $(0,
\cdots, 0, 1)$, respectively. According to \eqref{exchange} and \eqref{reverse}, $G_n^{0,0}$ equals to $(1,
0,\cdots, 0, 1)$ and $G_n^{0,1}$ equals to $(0, \cdots, 0)$, which
mean only exchange command is performed. This simple property guides us design algorithms with lower complexities.

\section{Encoding and Decoding Algorithms}
\label{sec-alg}

The encoding problems are studied first and four encoding algorithms are proposed. Two of them are based on
arithmetic operations and the other two bases on bit operations. Then decoding algorithms are obtained similarly.
The algorithms either have linear complexity or have constant complexity.

\subsection{Encoding Algorithms}
Let the level of Hilbert order be $m$. For any point, $(x_n, \cdots, x_1) (\in D_m)$,
each component $x_i (1 \le i \le n)$
is written as $x_i = (x_i^mx_i^{m-1}\cdots x_i^1)_2$. The calculated Hilbert order is
stored as $(r_mr_{m-1}\cdots r_1)_{2^n}$.

When calculating the Hilbert order, the reverse command is performed first, followed by
exchange command. The first encoding algorithm is described in
Algorithm \ref{alg-encoding1}. We assume $G_n^{i,0}$ and $G_n^{i,1}$ are known, which can be calculated
by algorithms in \cite{chenyang} and be stored.

\begin{algorithm}
\caption{Encoding algorithm}
\label{alg-encoding1}
\begin{algorithmic}
\State (1) If $m = 0$, terminate the procedure. Or we have $r_m =
f_n(x_n^m, x_{n-1}^m, \cdots, x_1^m)$.

\State (2) For each integer $i (1 \le i \le n)$, if $x_i^m$ equals to 1, then $x_i = x_i - 2^{m-1}$.

\State (3)\textit{Reverse}. For each integer $i (1 \le i \le n)$, if the $i$-th component of $G_n^{{r_m},1}$ is 1, then $x_i = 2^{m-1}
- 1 - x_i$.

\State (4)\textit{Exchange}. If $G_n^{{r_m},0}$ has two components equal to 1 in $i$-th and $j$-th position, then swap $x_i$ and
$x_j$.

\State (5) $m = m - 1$, goto (1).
\end{algorithmic}
\end{algorithm}

The Algorithm \ref{alg-encoding1} is an iterative algorithm, which loops from $m$ to $0$.
It's evident that for $n$ dimensional Hilbert order, the complexity of Algorithm \ref{alg-encoding1} is $O(nm)$. For
one specific dimension, such as two and three dimensional spaces, the complexity is $O(m)$, which is linear.

Now, let us analyze Algorithm \ref{alg-encoding1}. In step (2), the $i$-th component
$x_i$, $(x_i^m\cdots x_i^1)_2$, is replaced by
$(x_i^{m-1}\cdots x_i^1)_2$. This can be achieved by a simple $\&$ operation. And in
step (3), the reverse operation is equivalent to $Re_{m-1}$ operation.
Then algorithm \ref{alg-encoding1} can be rewritten with bit operations, which is described in Algorithm \ref{alg-encoding2}.

\begin{algorithm}
\caption{Encoding algorithm}
\label{alg-encoding2}
\begin{algorithmic}
\State (1) If $m = 0$, terminate the procedure. Or we have $r_m =
f_n(x_n^m, x_{n-1}^m, \cdots, x_1^m)$.

\State (2) For each integer $i (1 \le i \le n)$, $x_i =
(x_i^{m-1}\cdots x_i^1)_2 = (x_i^m\cdots x_i^1)_2 ~\&~ (011\cdots
1)_2$.

\State (3)\textit{Reverse}. For each integer $i (1 \le i \le n)$,
if the $i$-th component of $G_n^{{r_m},1}$ is 1, then $x_i =
Re_{m-1}(x_i)$.

\State (4)\textit{Exchange}. If $G_n^{{r_m},0}$ has two components,
which equal to 1 in $i$-th and $j$-th position, then swap $x_i$ and
$x_j$.

\State (5) $m = m - 1$, goto (1).

\end{algorithmic}
\end{algorithm}

Algorithm \ref{alg-encoding2} has the same complexity as Algorithm \ref{alg-encoding1}. The only difference
is that it uses bit operations other than arithmetic operations. The algorithm should be more efficient.

\subsection{Encoding Algorithms with Lower Complexities}
Now we develop encoding algorithms with lower complexities. According to our analysis above,
$G_n^{0,0}$ has two nonzero components,
which indicates where exchange command would be performed, and all components of $G_n^{0,1}$ are zero,
which means that no inverse
command would be performed. From our encoding algorithms above,
we can see that only exchange command between $x_n$ and $x_1$ is performed until value of $f_n$ isn't zero.
The $r_m$ is zero all the time. Therefore we can skip part of the loop.
This property shows us opportunity to reduce the number of iterations. Assume $log_2(0)$ equals to 0, we define $k =
floor(log_2(max\{x_n, \cdots, x_1\})) + 1$. Then the two encoding algorithms
above are rewritten to Algorithm \ref{alg-encoding3} and Algorithm \ref{alg-encoding4} respectively.

\begin{algorithm}[htb]
\caption{Encoding algorithm}
\label{alg-encoding3}
\begin{algorithmic}
\State (1) Set $(r_m\cdots r_1)_{2^n}$ to 0. If $m$ and $k$ have
different parities, then swap $x_1$ and $x_n$. $m = k$.

\State (2) If $m = 0$, terminate the procedure. Or we have $r_m =
f_n(x_n^m, x_{n-1}^m, \cdots, x_1^m)$.

\State (3) For each integer $i (1 \le i \le n)$, if $x_i^m$
equals to 1, then $x_i = x_i - 2^{m-1}$.

\State (4)\textit{Reverse}. For each integer $i (1 \le i \le n)$,
if the $i$-th component of $G_n^{{r_m},1}$ is 1, then $x_i = 2^{m-1}
- 1 - x_i$.

\State (5)\textit{Exchange}. If $G_n^{{r_m},0}$ has two components,
which equal to 1 in $i$-th and $j$-th position, then swap $x_i$ and
$x_j$.

\State (6) $m = m - 1$, goto (2).

\end{algorithmic}
\end{algorithm}

\begin{algorithm}[htb]
\caption{Encoding algorithm}
\label{alg-encoding4}
\begin{algorithmic}
\State (1) Set $(r_m\cdots r_1)_{2^n}$ to 0. If $m$ and $k$ have
different parities, then swap $x_1$ and $x_n$. $m = k$.

\State (2) If $m = 0$, terminate the procedure. Or we have $r_m =
f_n(x_n^m, x_{n-1}^m, \cdots, x_1^m)$.

\State (3) For each integer $i (1 \le i \le n)$, $x_i =
(x_i^{m-1}\cdots x_i^1)_2 = (x_i^m\cdots x_i^1)_2 ~\&~ (011\cdots
1)_2$.

\State (4)\textit{Reverse}. For each integer $i (1 \le i \le n)$,
if the $i$-th component of $G_n^{{r_m},1}$ is 1, then $x_i =
Re_{m-1}(x_i)$.

\State (5)\textit{Exchange}. If $G_n^{{r_m},0}$ has two components,
which equal to 1 in $i$-th and $j$-th position, then swap $x_i$ and
$x_j$.

\State (6) $m = m - 1$, goto (2).
\end{algorithmic}
\end{algorithm}

Again algorithm \ref{alg-encoding3} is based on arithmetic operations and Algorithm \ref{alg-encoding4} is
based on bit operations, which terminate in $k$ steps, thus the complexities
of them are $O(nk)$. For any given point $P$, the $k$ is fixed, therefore, the number of iterations
of the latter two algorithms are independent of the level
of Hilbert curve. For any specific dimension, the complexity is constant.
We can see that they are more efficient than the former two when level $m$ is much larger than $k$.

\subsection{Decoding Algorithms}

Let the level of Hilbert curve be $m$. For any integer $z \in [0, 2^{nm})$, it can be written as $(r_mr_{m-1}\cdots
r_1)_{2^n}$. Final result is a coordinate, $(x_n, \cdots, x_2, x_1) (\in D_m)$. Decoding procedure is the
reverse process of encoding procedure. In the decoding procedure, the exchange command is performed first,
then the reverse command. The first decoding algorithms is described in
Algorithm \ref{alg-decoding1}. Again, each arithmetic operation in Algorithm \ref{alg-decoding1}
is equivalent to a bit operation, and a decoding algorithm based on bit operations is proposed by the
Algorithm \ref{alg-decoding2}.

\begin{algorithm}
\caption{Decoding algorithm}
\label{alg-decoding1}
\begin{algorithmic}
\State (1) Let $(x_n, \cdots, x_2, x_1) = b_n(r_1)$ and $v = 2$.

\State (2) If $v > m$, terminate the procedure. Or we have
$b_n(r_v) = (s_n, \cdots, s_2, s_1)$.

\State (3)\textit{Exchange}. If $G_n^{{r_v},0}$ has two components,
which equal to 1 in $i$-th and $j$-th position, then swap $x_i$ and
$x_j$.

\State (4)\textit{Reverse}. For each integer $i (1 \le i \le n)$,
if the $i$-th component of $G_n^{{r_v},1}$ is 1, then $x_i = 2^{v-1}
- 1 - x_i$.

\State (5) For each integer $i (1 \le i \le n)$, if $s_i$ equals
to 1, then $x_i = x_i + 2^{v-1}$.

\State (6) $v = v + 1$, goto (2).
\end{algorithmic}
\end{algorithm}

\begin{algorithm}
\caption{Decoding algorithm}
\label{alg-decoding2}
\begin{algorithmic}
\State (1) Let $(x_n, \cdots, x_2, x_1) = b_n(r_1)$ and $v = 2$.

\State (2) If $v > m$, terminate the procedure. Or we have
$b_n(r_v) = (s_n, \cdots, s_2, s_1)$.

\State (3)\textit{Exchange}. If $G_n^{{r_v},0}$ has two components,
which equal to 1 in $i$-th and $j$-th position, then swap $x_i$ and
$x_j$.

\State (4)\textit{Reverse}. For each integer $i (1 \le i \le n)$,
if the $i$-th component of $G_n^{{r_v},1}$ is 1, then $x_i =
Re_{v-1}(x_i)$.

\State (5) For each integer $i (1 \le i \le n)$, if $s_i$ equals
to 1, then $x_i = (x_i) \wedge (1 \ll (v-1))$.

\State (6) $v = v + 1$, goto (2).
\end{algorithmic}
\end{algorithm}

Algorithm \ref{alg-decoding1} and Algorithm \ref{alg-decoding2} terminate in $m$ steps
and it's evident that their complexities are $O(nm)$. For any specific dimension $n$, the complexity
is linear.

\subsection{Decoding Algorithms with Lower Complexities}
We introduce the definition of $k$ here, which means $r_k > 0$ and $r_i = 0 (i > k)$. It is the location that
the first $r_i$ is not zero. We also assume $k$ equals to 1 if $z$ equals to 0.
It's equivalent to $k = floor(log_{2^n}((r_m\cdots r_1)_{2^n})) + 1$ if we assume $log_{2^n}(0)$ equals to 0. $k$ isn't greater
than $m$ and is independent of $m$. Now we can rewrite Algorithm \ref{alg-decoding1} and Algorithm \ref{alg-decoding2}
with lower complexities.

\begin{algorithm}
\caption{Decoding algorithm}
\label{alg-decoding3}
\begin{algorithmic}
\State (1) Let $(x_n, \cdots, x_2, x_1) = b_n(r_1)$ and $v = 2$.

\State (2) If $v > k$, terminate the procedure. Or we have
$b_n(r_v) = (s_n, \cdots, s_2, s_1)$.

\State (3)\textit{Exchange}. If $G_n^{{r_v},0}$ has two components,
which equal to 1 in $i$-th and $j$-th position, then swap $x_i$ and
$x_j$.

\State (4)\textit{Reverse}. For each integer $i (1 \le i \le n)$,
if the $i$-th component of $G_n^{{r_v},1}$ is 1, then $x_i = 2^{v-1}
- 1 - x_i$.

\State (5) For each integer $i (1 \le i \le n)$, if $s_i$ equals
to 1, then $x_i = x_i + 2^{v-1}$.

\State (6) $v = v + 1$, goto (2).

\State (7) If $m$ and $k$ have different parities, then swap $x_n$ and $x_1$.
\end{algorithmic}
\end{algorithm}

\begin{algorithm}
\caption{Decoding algorithm}
\label{alg-decoding4}
\begin{algorithmic}
\State (1) Let $(x_n, \cdots, x_2, x_1) = b_n(r_1)$ and $v = 2$.

\State (2) If $v > k$, terminate the procedure. Or we have
$b_n(r_v) = (s_n, \cdots, s_2, s_1)$.

\State (3)\textit{Exchange}. If $G_n^{{r_v},0}$ has two components,
which equal to 1 in $i$-th and $j$-th position, then swap $x_i$ and
$x_j$.

\State (4)\textit{Reverse}. For each integer $i (1 \le i \le n)$,
if the $i$-th component of $G_n^{{r_v},1}$ is 1, then $x_i =
Re_{v-1}(x_i)$.

\State (5) For each integer $i (1 \le i \le n)$, if $s_i$ equals
to 1, then $x_i = (x_i) \wedge (1 \ll (v-1))$.

\State (6) $v = v + 1$, goto (2).

\State (7) If $m$ and $k$ have different parities, then swap
$x_n$ and $x_1$.
\end{algorithmic}
\end{algorithm}

Algorithm \ref{alg-decoding3} and Algorithm \ref{alg-decoding4} terminate in $k$ steps and
therefore the complexities are $O(nk)$. For any fixed input and any specific dimension, the complexity
is constant.

\section{Discussion}
\label{sec-dis}
Encoding and decoding algorithms above are concise and easy to implement.
In this section, a two dimensional example is
employed to illustrate how to use these algorithms. For the sake of simpleness, only
Algorithm \ref{alg-encoding1} is studied.

When $n$ equals to two, the functions $b_2$ and $f_2$ are calculated and stored in
Table~\ref{map}. Exchange command $G_2^{i, 0}$ and reverse command $G_2^{i, 1}$ are listed in Table~\ref{gene},
hich are obtained using algorithms introduced in \cite{chenyang}.

\begin{table}
\centering
\caption{Function $f_2$ and $b_2$.}
{
  \begin{tabular}{|c|c|c|c|} \hline
   $f_2(0, 0)$& 0 &  $b_2(0)$& (0, 0)\\
   $f_2(0, 1)$& 1 &  $b_2(1)$& (0, 1)\\
   $f_2(1, 1)$& 2 &  $b_2(2)$& (1, 1)\\
   $f_2(1, 0)$& 3 &  $b_2(3)$& (1, 0)\\ \hline
  \end{tabular}
}
\label{map}
\end{table}

\begin{table}
\centering
\caption{Exchange commands and reverse commands.}
{
  \begin{tabular}{|c|c|c|c|} \hline
   $G_2^{0,0}$& (1, 1) &  $G_2^{0,1}$& (0, 0)\\
   $G_2^{1,0}$& (0, 0) &  $G_2^{1,1}$& (0, 0)\\
   $G_2^{2,0}$& (0, 0) &  $G_2^{2,1}$& (0, 0)\\
   $G_2^{3,0}$& (1, 1) &  $G_2^{3,1}$& (1, 1)\\ \hline
  \end{tabular}
}
\label{gene}
\end{table}

Checking Algorithm \ref{alg-encoding1}, we can see what steps (2) $\sim$ (4) do is to update coordinate $(x_n,\cdots, x_1)$ only.
The reason we describe the process by using three steps is to make the process clear. In practice, these three steps can be combined
together and we use one step to update directly. The updating rules for encoding procedure are shown in Table~\ref{encoding}, where $(x_2,
x_1)$ is replaced by $(x, y)$.

The updating rules can be obtained easily using our algorithm. Take $i$
equals to 3 as an example. In this case, $G_2^{3,0}$ is $(1,1)$, and $G_2^{3,1}$ is (1, 1). According to our analysis,
an exchange operation and two reverse operations should be performed. The new coordinate $(x_{new}, y_{new})$ is
$(x - 2^{m-1}, y)$ after step (2). It is $(2^m -1 -x, 2^{m-1}-1-y)$ after reverse operation. Then we perform the exchange command, the final coordinate is $(2^{m-1}-1-y, 2^m -1 -x)$, which is the same as shown in Table~\ref{encoding}. Updating rules for decoding algorithm
can be obtained similarly.

\begin{table}
\centering
\caption{Updating rules  of encoding procedure.}
{
\begin{tabular}{|c|c|c|} \hline
   Quadrant &  $x_{new}$ & $y_{new}$ \\ \hline
   $0$& $y$ & $x$ \\
   $1$& $x$ & $y - 2^{m-1}$ \\
   $2$& $x - 2^{m-1}$& $y - 2^{m-1}$\\
   $3$& $2^{m-1} - 1 - y$ & $2^m - 1 - x$ \\ \hline
\end{tabular}
}
\label{encoding}
\end{table}

In \cite{ningtao}, Chen et al. developed a reduced complexity algorithm for two dimensional case
and numerical results were also presented. We implement the three dimensional case in PHG \cite{phg,hliu},
which is an essential module for dynamic load balancing. The Hilbert order algorithms were also implemented
in Zoltan \cite{Zoltan}. The performance data is collected in Table \ref{en-perf}.
The first row represents the level of Hilbert order, and others represent running time.
>From this table, we can see that for a fixed coordinate (1, 1, 1), the computational time increases linearly for
algorithm with the $O(m)$ complexity while it's fixed for algorithm with $O(k)$ complexity.

\begin{table}
\centering
\caption{Performance of encoding algorithm for (1,1,1)}
{
\begin{tabular}{|c|c|c|c|c|} \hline
   Algorithm & 8 (s)    & 32 (s)  & 128 (s) & 256 (s)   \\ \hline
   O(m)      & 1.60E-7  & 4.69E-7 & 1.67E-6 & 3.25E-6   \\
   O(k)      & 8.38E-8  & 8.39E-8 & 8.39E-8 & 8.38E-8   \\ \hline
\end{tabular}
}
\label{en-perf}
\end{table}

\section{CONCLUSION}

Encoding and decoding of arbitrary dimension Hilbert order are studied. And the open problem is solved by
this paper. Four encoding algorithms and four decoding algorithms are proposed. Four of them have linear
computation complexities and the other four have lower complexities.
The properties of evolutive rules introduced by \cite{chenyang} are also studied and some results are deduced,
which are applied to develop encoding and decoding algorithms.
A two dimensional case is studied and updating rules are presented.
By using algorithms developed
in this paper, algorithms for any specific dimensional spaces can be obtained.
In the end of the paper,
numerical experiments are performed, which demonstrate the difference between algorithms with difference
complexities.

\section*{Acknowledgments}
This work is supported by the 973 Program under the grant
 2011CB309703, by China NSF under the grants 11021101 and 11171334, by the 973 Program under
 the grant 2011CB309701, the China NSF under the grants 11101417
and by the National Magnetic Confinement Fusion Science Program under the grants 2011GB105003.

\end{document}